\documentclass{jetpl}
\twocolumn

\lat

\title{Non-exponential decoherence of radio-frequency resonance rotation of spin in storage rings}

\rtitle{Non-exponential decoherence\ldots}

\sodtitle{Non-exponential decoherence of radio-freqency resonance rotation of spin in storage rings }

\author{A.\,Saleev$^{1+}$,
N.\,N.\,Nikolaev$^{2*}$, F.\,Rathmann$^{1}$, F.\,Hinder$^{1=}$, J.\,Pretz$^{1=3}$, M.\,Rosenthal$^{=4}$}
\rauthor{A.\,Saleev, 	N.\,N.\,Nikolaev, F.\,Rathmann, F.\,Hinder, J.\,Pretz, M.\,Rosenthal}

\sodauthor{Saleev, 	Nikolaev, Rathmann, Hinder, Pretz, Rosenthal}

\address{
$^1$Institut f\"ur Kernphysik, Forschungszentrum J\"ulich, 52425 J\"ulich, \\~\\
$^+$Samara National Research University, 443086 Samara,  \\~\\
$^2$L.D.Landau Institute for Theoretical Physics RAS, 117940 Moscow, Russia,\\~\\
$^*$Moscow Institute for Physics and Technology, 141700 Dolgoprudny,  \\~\\
$^=$III. Physikalisches Institut B, RWTH Aachen University, 52056 Aachen, Germany, \\~\\
$^3$JARA-FAME (Forces and Matter Experiments). Forschungszentrum J\"ulich and RWTH Aachen University, Germany, \\~\\
$^4$CERN, Route de Meyrin
Meyrin 1217, Switezerland.}
\abstract{Precision experiments, such as the search for electric dipole moments of charged particles using radiofrequency spin rotators in storage rings, demand for maintaining the exact spin resonance  condition for several thousand seconds. Synchrotron oscillations in the stored beam modulate the spin tune of off-central particles, moving it off the perfect resonance condition set for central particles on the reference orbit. Here we report an analytic description of  how synchrotron oscillations lead to non-exponential decoherence of the radiofrequency resonance driven up-down  spin rotations. This non-exponential decoherence is shown to be accompanied by a nontrivial walk of the spin phase. We also comment on sensitivity of the decoherence rate to the harmonics of the radiofreqency spin rotator and a possibility to check predictions of decoherence-free magic energies. }


\begin{document}		
\maketitle
		
	%\tableofcontents
	%%%%%%%%%%%%%%%%%%%%%%%%%%%%%%%%%%%%%%%%%%%%%%%%%%%%%%%%%%%%%%%%%%%%%%%%%%%%%%%%
	%%%%%%%%%%%%%%%%%%%%%%%%%%%%%%%  Section I %%%%%%%%%%%%%%%%%%%%%%%%%%%%%%%%%%%%%
	%%%%%%%%%%%%%%%%%%%%%%%%%%%%%%%%%%%%%%%%%%%%%%%%%%%%%%%%%%%%%%%%%%%%%%%%%%%%%%%%
	\section{Introduction}
	%%%%%%%%%%%%%%%%%%%%%%%%%%%%%%%%%%%%%%%%%%%%%%%%%%%%%%%%%%%%%%%%%%%%%%%%%%%%%%%%
	Radiofrequency (RF) spin rotator driven resonance up-down  oscillations  of polarization of charged particles in storage rings are of interest in a broad class of spin experiments incuding the search for electric dipole moments of charged particles \cite{JEDI,Anastassopoulos:2015ura,Orlov:2006su}. Inherent to particles in storage rings is a modulation of the spin-tune by  synchrotron oscillations (SO) which drives the spin precession and RF frequencies apart \cite{Benati:2012pv} and decoheres evolution of the polarization of particles in an ensemle. It becomes still more important at large spin rotation times imperative in searches for extremely small EDM's \cite{Rathmann:2013rqa,SCT,Hempelmann:2017zgg}. In this paper we develop a fully analytic theory of SO driven spin decoherence and  show that SOs entail a non-exponential attenuation of resonance driven up-down oscillations averaged over an ensemle. We also derive the associated nontrivial spin phase walk and the dependence of non-exponential decoherence on the harmonics of the RF spin rotator. These analytic results are new and do substantially extend the early considerations by Benati et al. \cite{Benati:2012pv}
	
    SO driven decoherence depends on the spread of SO amplitudes $a_z$ in an ensemble. Hence first we relate a distribution  of SO amplitudes,  $F(a_z)$, to longitudinal density $N(z)$ in the bunch.
	A particle with SO amplitude $a_z$ oscillates around the center of bunch,
	$z=a_z \cos(2\pi \nu_z f_R t +\lambda) $, where $f_R$ is the ring frequency,  $\nu_z$ is  the synchrotron tune,  $\lambda \in [0,2\pi]$ is random phase and we only need to consider  $a_z> 0$. Evidently, the one-particle density $N(z) \propto 1/\Delta \beta_z$, where $
	\Delta \beta_z c= 2\pi \nu_z f_R (a_z^2 -z^2)^{1/2}$
	 is a velocity of synchrotron oscillations, c is velocity of light. For an ensemble 
	\begin{equation}
	N(z) = \frac{1}{\pi} \int_z^\infty \frac{da_z F(a_z)}{\sqrt{a_z^2 -z^2} }\, \label{eq:Nz}
	\end{equation}
	and has a form of the Abel transform with solution \cite{Tricomi}
	\begin{equation}
	F(a_z)= -2a_z \int_{a_z}^\infty  \frac{dz N'(z)}{\sqrt{z^2-a_z^2}} \, . \label{eq:Faz}
	\end{equation}
	
	With Gaussian $N(z) \propto \exp(-z^2/2B^2)$, suggested by the experimental studies \cite{Marcel} for the sinusoidal potential RF-cavity, a solution for $F(a_z)$ is readily obtained in the closed analytic form:
	\begin{align}
	F(a_z)&=\frac{a_z}{B^2}\exp\left(-\frac{a_z^2}{2B^2}\right) \nonumber\\
	&\approx \frac{1}{\sqrt{\pi}B} \exp\left(-\frac{(a_z-B)^2}{B^2}\right)\,. \label{eq:analytic}
	\end{align}
	We can relate $B$ to the momentum spread in the bunch
	\begin{equation} 
	B = \frac{c}{f_R} \cdot \frac{\beta}{\gamma^2 \pi \sqrt{3}}\cdot\frac{1}{\nu_z}\Bigl\langle \frac {\Delta p ^2}{p^2} \Bigr\rangle^{1/2}\,, \label{eq:bunchlength}
	\end{equation}
	where $\beta c$ and $\gamma$ are the velocity and gamma-factor of stored particle. 
	
    Now we treat the one particle-problem of a spin motion under continuous operation of the RF spin rotator. SOs modulate the  spin tune $\nu_s$ of off-central stored particles, $\Delta \nu_s = G\Delta\gamma$. The  spin phase after n revolutions will be
	$\theta_s(n) = 2\pi \nu_s n + \Delta \theta_s(n)$, where the spin phase slip
	\begin{align}
	\Delta\theta_s(n)
	 = \psi_s \xi [\cos(2\pi \nu_z n+\lambda)-\cos\lambda]  \label{eq:spinphaseshift}
	\end{align}
	and $\xi=a_z/B$. Typically the spin phase slip parameter 
	\begin{equation}
	\psi_s = \frac{2\pi G\beta \gamma^3 f_R B}{c} \ll 1\,, \label{eq:psi}
	\end{equation}
	especially for deuterons with small magnetic anomaly $G=(g-2)/2$. 
	
	A spin kick per revolution depends on the  phase of the RF field at exactly when a particle passes the spin rotator. SOs modulate the revolution  time $\tau$,
	\begin{equation}
	f_R  \Delta \tau=\frac{\eta}{\gamma\beta^2} \Delta \gamma \, ,
	\end{equation}
	where 
	$\eta$ 
	is the phase slip factor. This $\Delta \tau$ changes a time $t(n)$ of arrival of a particle at the RF spin rotator,
	\begin{equation}
	t(n) = \frac{n}{f_R} + \sum_{k=1}^n \Delta \tau_k\, .
	\end{equation} 
	Consequently, a phase of the RF spin rotator,  $\theta_{rf}(n)$, acquires the slip term,  $\Delta\theta_{rf}(n)$, 
	\begin{align}
	\theta_{rf}(n) &= 2\pi f_{rf} t(n)=2\pi (\nu_{rf}+K)n + \Delta\theta_{rf}(n) \nonumber\\
	&= 2\pi (\nu_{rf}+K)n + \frac{\nu_{rf}+K}{\nu_s}\cdot \frac{\eta}{\beta^2}\Delta\theta_s(n)\nonumber\\
	&= 2\pi (\nu_{rf}+K)n + (C_{rf} +1)\Delta\theta_s(n)\, ,
	\end{align}
	where $K=0,\pm 1, \pm 2...$ is the harmonics number in the RF frequency $f_{rf}=(\nu_{rf}+K)f_R$. Take note of correlated spin and RF phase slips \cite{LLMNR}.

	In further derivation of the impact of SO on the RF driven rotation of spin of stored particles, we follow closely the formalism of Appendix A in Ref. \cite{Mapping}. 
	RF spin rotator is described by the spin transfer matrix
	\begin{equation}
	{t}_{rf}(k)  =  \cos\frac{1}{2}\chi_{rf}(k) - i (\vec{\sigma} \cdot \vec{w} )\sin\frac{1}{2}\chi_{rf}(k) \, .  
	\label{eq:App-B.3}
	\end{equation}
	where $\vec{w}$ is the spin rotation axis, $\vec{\sigma}$ are the Pauli matrices and the spin kick for revolution k equals $\chi_{rf}(k) = \chi_{rf} \cos\left(\theta_{rf}(k) \right)$.

	Evolution of the spinor wave function $\phi$ of the stored particle per revolution $k$ is described by spin transfer matrices of the ring, ${T}_k$, and the RF rotator, \mbox{$
	\phi(k+1) =  {t}_{rf}(k){T}_k\phi(k)\,$}.  We pass to the conventional interaction representation with  ${T}(n)=\prod_{k=1}^n {T}_k $,
	\begin{align}
	\phi(n) &= {T}(n) \zeta(n) \nonumber\\
	&= 
	\left\{\cos\frac{1}{2}\theta_s(n) - i (\vec{\sigma} \cdot \vec{c} )\sin\frac{1}{2}\theta_s(n)\right\}\zeta(n)  \, .  
	\end{align}
	 Here $\vec{c}$ is the stable spin axis  and  the spinor $\zeta(n)$ describes the envelope over the rapid oscillations of the spin, $\zeta(0)=\phi(0)$. The evolution equation for $\zeta(n)$ 
	 becomes
	\begin{align}
	\zeta(n) &=  T^{-1}(n){t}_{rf}(n) T(n) \zeta(n-1)\nonumber\\
	& = \exp \left\{-  \frac{i}{2} \vec{\sigma} \cdot \vec{U}(n) \right\}\zeta(n-1)\, . \label{eq:evolutionequation}
	\end{align}
	Here 
	\begin{align}
		\vec U (n) &=   2 \sin \left( \frac{1}{2} \chi_{rf}(n)\right) 
	\times \Bigl\{\cos\theta_s(n)[[\vec c \times \vec w]\times \vec c]  \nonumber\\
	& -  \sin\theta_s(n)[\vec c \times \vec w] + (\vec c \cdot \vec w)\vec c \Bigr\}
	\label{eq:App-B.6}
	\end{align}
	is the spin rotation axis which lies in the plane rotating with frequency locked to the spin precession frequency, for details see  \cite{Mapping}.

	Equation Eq.\,(\ref{eq:evolutionequation}) has a  solution in terms of the n-ordered exponential
	\begin{equation}
	\zeta(n) = T_n \exp \left\{-  \frac{i}{2} \sum_{k=1}^n \vec{\sigma} \cdot \vec{U}(k) \right\}\psi(0) \, .	
	\label{eq:App-B.5}
	\end{equation}
	Hereafter we work to the lowest order in a small parameter $\chi_{rf} \ll 1$. For central particles the resonance condition is $\nu_{rf}=\nu_s$ and $f_{rf} = (\nu_s + K)f_{\text{R}} $, where the integer $K=0, \pm 1, \pm 2,... $ is the harmonics number.  The large-$n$ behavior of $\zeta(n)$ is evaluated using the Bogolyubov-Krylov-Mitropolsky averaging method\,\cite{Bogolyubov}. It amounts to keeping in the sum $\sum_{k=1}^n \vec{\sigma} \cdot \vec{U}(k)$  only the linearly rising terms
	\begin{align}
	&\sum_{k=1}^{n} 2 \chi_{rf}(k) \cos \theta_s(k)  = \chi_{rf}\sum_{k=1}^{n} 
	\cos \left[\theta_{\text{rf}}(k) - \theta_s(k)\right]\nonumber \\
	&=\chi_{ rf}\sum_{k=1}^{n} \cos \left\{C_{rf}\psi_s \xi [\cos(2\pi \nu_z n+\lambda)-\cos\lambda]\right\}\nonumber\\
	&=n\chi_{rf}\cos \left(C_{ rf}\psi_s \xi \cos\lambda\right) J_0\left(C_{rf}\psi_s \xi\right)\,,\\
		&\sum_{k=1}^{n} 2 \chi_{rf}(k) \sin \theta_s(k)  =\nonumber\\ 
			&=-n\chi_{rf}\sin \left(C_{ rf}\psi_s \xi \cos\lambda\right) J_0\left(C_{rf}\psi_s \xi\right)\,,
		\label{eq:App-B.8}
	\end{align}
	where $J_0(x)$ is the Bessel function and, as derived in \cite{LLMNR},
	\begin{equation}
	C_{rf}= \frac{\eta}{\beta^2}\cdot\left(1+\frac{K}{G\gamma}\right)-1\, .
	\end{equation}
	
		The SO corrected large-n spin evolution law becomes
	\begin{equation}
	\zeta(n) = \exp\left\{-\frac{i}{2} n \epsilon(\xi) {\vec \sigma}\cdot \vec{u}\right\}\phi(0)\,,
	\label{eq:App-B.7}
	\end{equation} 
	where $\epsilon(\xi)=\epsilon_0  J_0\left(C_{rf}\psi_s \xi\right)$ and 
		\begin{equation} 
	\epsilon_0 = \frac{1}{2} \chi_{rf}[1- (\vec{c} \cdot \vec{w})^2]^{1/2}\,,
	\end{equation}
		is a spin resonance strength at vanishing SO's,  when polarization  along the stable spin axis $\vec{c}$ oscillates $\propto \cos(\epsilon_0 n)$. The spin rotation axis $\vec{u}$ in Eq.\,(\ref{eq:App-B.7}) equals
	\begin{align}
	\vec{u} =\frac{\cos\Delta_{rf}[[\vec c \times \vec w]\times \vec c]  +\sin \Delta_{rf}[\vec c \times \vec w]}
	{\sqrt{1- (\vec{c} \cdot \vec{w})^2}}
	\end{align} 
    and 
	$\Delta_{rf} = C_{ rf}\psi_s \xi \cos\lambda$. Hereafter the  up-down oscillations are defined with respect to the  closed spin orbit vector $\vec c$. 	The above derived analytic factor  $ J_0\left(C_{rf}\psi_s \xi\right)$  describes a SO driven spread of the spin resonance strength and spin phase of an individual particle over an ensemble of particles in a bunch.
	
	A final step is averaging of the up-down oscillations of one-particle polarizations over  an ensemble. We define the ensemble averaged polarization $A(n)$ as $\left( \vec{S}(n)\cdot\vec{c}\right)= \left( \vec{S}(0)\cdot\vec{c}\right) A(n)$. It does not depend on the SO phase $\lambda$. Upon averaging polarizations of individual particles  over the SO amplitudes with the distribution of Eq.\,(\ref{eq:analytic}), we obtain our main result: 
	\begin{align}
	&A(n) =
	\Re \langle 
	\exp\left\{-i n \epsilon(\xi) \right\}
	\rangle_{\xi}
	\nonumber \\
	&\simeq \Re
	\left\langle
	\exp\left\{-i n \epsilon_0 \left[1-\frac{1}{4} C_{rf}^2\psi_s^2 \xi^2 \right]\right\} 
	\right\rangle_{\xi}\nonumber\\
		&=(1-i\rho n )^{-1/2} \exp\left\{-i\epsilon_0 n + \frac{i\rho n}{1-i\rho n}\right\}\nonumber\\
	&=\frac{\exp \left\{- \sin^2\varphi(n)\right\} }{(1+\rho^2n^2)^{1/4}}\cos \left\{\epsilon_0 n -\kappa(n)\right\} \,, \label{eq:final}
 	\end{align}
	where 
	\begin{align}
	&\rho = \frac{1}{4} \epsilon_0 C_{rf}^2 \psi_s^2\, ,\\
	&\varphi(n)=\arctan(\rho n)\,, \\
	&\kappa(n)=\frac{1}{2} \left[ \varphi(n) +  \sin 2 \varphi(n)\right]\,.
	\end{align}
	In the above averaging over the SO amplitudes over the  distribution given by Eq.\,(\ref{eq:analytic}), we resorted to a steepest descent approximation.

	One comment on the above derivation is in order. In generic case, the synchrotron tune $\nu_z$ depends on the synchrotron amplitude. We observe that $\nu_z$ only enters the principal spin phase slip through the parameter $\psi_s$ of Eq.\,(\ref{eq:psi}) in the combination $\xi/\nu_z$. Consequently, weak dependence of $\nu_z$ on $\xi$ only would mimic a slight modification of the synchrotron amplitude distribution and would  affect none of our principal conclusions.  
	\begin{figure}[htb]
		\includegraphics[width=\columnwidth]{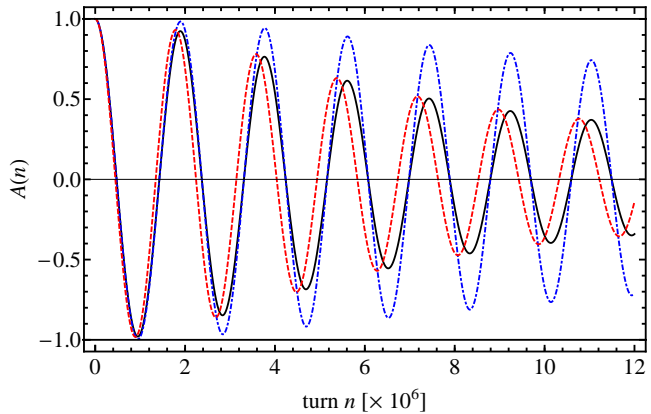}
		\caption{\label{fig:damping} Fig.1. Non-exponential attenuation of vertical polarization oscillations: the solid curve is given Eq.\,(\ref{eq:final}), the dashed (red) curve is Eq.\,(\ref{eq:final}) where we put spin phase walk $\kappa(n)=0$ and the dot-dashed (blue) curve is for suppressed attenuation factor $\exp \left\{- \sin^2\varphi(n)\right\}$.}
	\end{figure}
		
    Now we turn to discussion of our principal findings. Synchrotron oscillations do clearly decohere the RF driven resonance up-down oscillations of polarization of stored particles \cite{Benati:2012pv}.  The last line of Eq.\,(\ref{eq:final}) is  our final analytic result for the SO caused decohernce of RF driven up-down spin oscillations in storage rings. Typical decoherence pattern is shown by solid line in Fig.\,\ref{fig:damping}. Here we evaluated $A(n)$ for deuterons of momentum $p=970$ MeV/c, $\eta=-0.61$ \cite{Benati:2012pv}, the momentum spread 
	$\langle \Delta p^2 /p^2 \rangle ^{1/2}=3\cdot 10^{-4}$ and the RF driven up-down spin oscillations with the period $\tau_s = 2.4$s. This corresponds to RF solenoid with $\int B dl=0.0264$ T\,mm. Then in this example we have $\rho= 1.37\cdot 10^{-7}$.

	The overall attenuation of the envelope of the up-down spin oscillations comes from two non-exponential factors.  The first one, $ \exp \left\{- \sin^2\varphi(n)\right\}$ tends to a constant $1/e$ as soon as $\varphi(n)> 1$. A significance of this attenuation factor can be judged from a comparison of the solid line with the dot-dashed (blue) line  in Fig.\,\ref{fig:damping} - in the latter case we took out the factor  $ \exp \left\{- \sin^2\varphi(n)\right\}$. As such, the dot-dashed (blue) line in Fig.\,\ref{fig:damping} shows the effect of the second attenuation factor, $(1+\rho^2 n^2)^{-1/4}$, which decreases continuously $\propto 1/\sqrt{n}$. Attempts to describe this decoherence by conventional exponential attenuation of the oscillation envelope will be entirely misleading. 
	
\begin{figure}[htb]
		\includegraphics[width=\columnwidth]{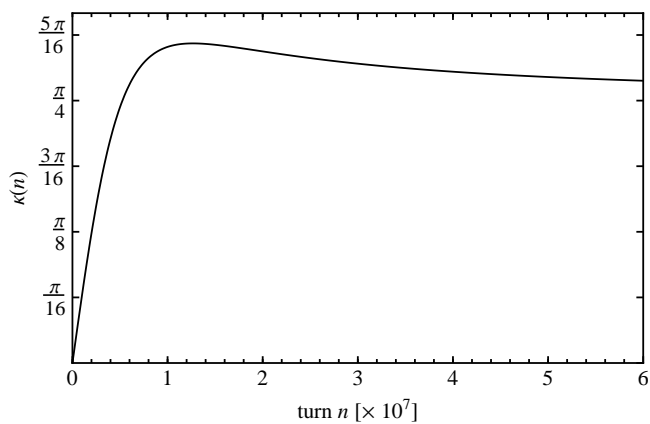}
		\caption{\label{fig:phase} Fig.2. Phase walk $\kappa(n)$ of vertical polarization oscillations. }
	\end{figure}

	Besides the non-exponential damping of spin oscillations, we predict a nontrivial walk of the spin phase $\kappa(n)$. This phase walk, shown in Fig.\,\ref{fig:phase} is clearly seen from a comparison of the solid and dashed (red) curves in Fig.\,\ref{fig:damping}, where the dashed curve (red) shows plane $\cos (\epsilon_0 n)$ oscillation law omitting the phase walk $\kappa(n)$. We predict a large-n limiting phase $\kappa(n) \to \pi/4$.
	
	Our predictions can be tested experimentally.  Our point is that the spin phase motion and attenuation of the spin envelope only depend on $\rho$. One needs to fit the up-down spin oscillations to our form of $A(n)$, which only depends on two free parameters: $\epsilon_0$ and $\rho$.
	
	We emphasize that $\rho$ must follow a generic pattern of dependence on the harmonics $K$ as it was predicted in Ref. \cite{LLMNR}, including the decoherence-free magic energies at which $C_{rf}\approx 0$ - it would be especially interesting to check experimentally these predictions. 
	
	The factor $C_{rf}$ is readily calculable in terms of the known slip-factor $\eta$, and one can readily convert the fitted $\rho$ into the r.m.s. bunch length parameter $B$. On the other hand, one can determine $B$ experimentally from the time distribution of interactions in the internal target \cite{Marcel}, what would offer important crosscheck of our analytic results. All these tests, including the dependence on the synchrotron tune and on the momentum spread in the beam, both only entering via the bunch length, see Eq.\,(\ref{eq:bunchlength}), can be performed in future experiments at COSY.
	
	In this short communication we restricted ourselves to analytic results only for up-down oscillations of polarization for the resonance case, $\nu_{rf}=\nu_s$. Following the technique exposed in Appendix A of Ref. \cite{Mapping}, one can redaily extend the above considerations to the off-resonance case. Of particular interest will be a decoherence of polarization in the ring plane.  Also of practical interest will be  side band resonances at $\nu_{rf}=\nu_s \pm \nu_z$.. These extensions and detailed numerical simulations of dependences on the synchrotrone tune and the beam momentum spread, based on the orbit and spin trackers, will be reported elsewhere.
	
	This study was a part of the JEDI Collaboration efforts in spin dynamics aimed at future searches for EDMs of charged particles at storage rings \cite{JEDI}, supported by an ERC Advanced Grant (srEDM $\#$ 694390). We are grateful to JEDI members for useful discussions and suggestions.
	The work of N.N.\ Nikolaev and A.\,Saleev was supported by a Grant from the Russian Science Foundation (grant number RNF-16-12-10151).


\begin{thebibliography}{99}

%1
\bibitem{JEDI}
JEDI Collaboration, proposals available from http://collaborations.fz-juelich.de/ikp/jedi/

%2
%\cite{Anastassopoulos:2015ura}
\bibitem{Anastassopoulos:2015ura} 
V.~Anastassopoulos {\it et al.},
%``A Storage Ring Experiment to Detect a Proton Electric Dipole Moment,''
Rev.\ Sci.\ Instrum.\  {\bf 87}, no. 11, 115116 (2016)
doi:10.1063/1.4967465
[arXiv:1502.04317 [physics.acc-ph]].
%%CITATION = doi:10.1063/1.4967465;%%
%28 citations counted in INSPIRE as of 02 Jul 2017


%3
%\cite{Orlov:2006su}
\bibitem{Orlov:2006su} 
Y.~F.~Orlov, W.~M.~Morse and Y.~K.~Semertzidis,
%``Resonance method of electric-dipole-moment measurements in storage rings,''
Phys.\ Rev.\ Lett.\  {\bf 96}, 214802 (2006)
doi:10.1103/PhysRevLett.96.214802
[hep-ex/0605022].
%%CITATION = doi:10.1103/PhysRevLett.96.214802;%%
%71 citations counted in INSPIRE as of 02 Jul 2017

%4	
%\cite{Benati}
\bibitem{Benati:2012pv}
P.~Benati {\it et al.},
%``Synchrotron oscillation effects on an rf-solenoid spin resonance,''
Phys.\ Rev.\ ST Accel.\ Beams {\bf 15} (2012) 124202
Erratum: [Phys.\ Rev.\ ST Accel.\ Beams {\bf 16} (2013) no.4,  049901].
doi:10.1103/PhysRevSTAB.15.124202, 10.1103/PhysRevSTAB.16.049901
%%CITATION = doi:10.1103/PhysRevSTAB.15.124202, 10.1103/PhysRevSTAB.16.049901;%%
%12 citations counted in INSPIRE as of 02 Jul 2017
%\cite{Eversmann:2015jnk}

%5
\bibitem{Rathmann:2013rqa}
F.\, Rathmann, A.\,Saleev and N.\,Nikolaev
The search for electric dipole moments of light ions in
storage rings,
J. Phys. Conf. Ser.,
{\bf 447}, 012011 (2013); 

%6
%\cite{Guidoboni:2016bdn}
\bibitem{SCT} 
G.~Guidoboni {\it et al.} [JEDI Collaboration],
%``How to Reach a Thousand-Second in-Plane Polarization Lifetime with 0.97-GeV/c Deuterons in a Storage Ring,''
Phys.\ Rev.\ Lett.\  {\bf 117}, no. 5, 054801 (2016).
doi:10.1103/PhysRevLett.117.054801
%%CITATION = doi:10.1103/PhysRevLett.117.054801;%%
%6 citations counted in INSPIRE as of 02 Jul 2017

%7

\bibitem{Hempelmann:2017zgg}
N.~Hempelmann {\it et al.} [JEDI Collaboration],
%``Phase locking the spin precession in a storage ring,''
Phys.\ Rev.\ Lett.\  {\bf 119} (2017) no.1,  014801
doi:10.1103/PhysRevLett.119.014801
[arXiv:1703.07561 [physics.acc-ph]].

%8
\bibitem{Tricomi}
F.\,G.\,Tricomi, Integral Equations (Pure and applied mathematics, v. 5), (1985), Dover Publications.


%9
\bibitem{Marcel}
M.\,S.\,Rosenthal, PhD Thesis, 
Experimental Benchmarking of Spin Tracking Algorithms
for Electric Dipole Moment Searches at the Cooler
Synchrotron COSY, RWTH Aachen U. (2016),
http://publications.rwth-aachen.de/record/671012



%10
%\cite{Lehrach:2012eg}
\bibitem{LLMNR} 
A.~Lehrach, B.~Lorentz, W.~Morse, N.~Nikolaev and F.~Rathmann,
%``Precursor Experiments to Search for Permanent Electric Dipole Moments (EDMs) of Protons and Deuterons at COSY,''
arXiv:1201.5773 [hep-ex].
%%CITATION = ARXIV:1201.5773;%%
%25 citations counted in INSPIRE as of 02 Jul 2017

%11
%\cite{Mapping}
\bibitem{Mapping} 
A.~Saleev {\it et al.} [JEDI Collaboration],
%``Spin tune mapping as a novel tool to probe the spin dynamics in storage rings,''
Phys.\ Rev.\ Accel.\ Beams {\bf 20} (2017) no.7,  072801
[Phys.\ Rev.\ Accel.\ Beams.\  {\bf 20} (2017) 072801]
doi:10.1103/PhysRevAccelBeams.20.072801
[arXiv:1703.01295 [physics.acc-ph]].
%%CITATION = doi:10.1103/PhysRevAccelBeams.20.072801;%%

%12
\bibitem{Bogolyubov}
N.\,N.\,Bogoliubov and Y.\,A.\,Mitropolsky,
Asymptotic Methods in the Theory of Non-linear Oscillation,
Gordon and Breach, New York  (1961).



\end{thebibliography}
\end{document}